\documentclass[11pt]{article}
\usepackage{setspace}

\def\be{\begin{equation}}

\def\ee{\end{equation}}

\newcommand\of[1]{\left( #1 \right)}
\newcommand\sqof[1]{\left[ #1 \right]}

\def\bea{\begin{eqnarray}}

\def\eea{\end{eqnarray}}

\newcommand\refeq[1]{(\ref{#1})}
\begin{document}

\singlespace

\begin{flushright} BRX TH-617 \\
CALT 68-2770
\end{flushright}

\vspace*{.3in}

\begin{center}

{\Large\bf  De/re-constructing the Kerr Metric}

{\large S.\ Deser$^1$ and J.\ Franklin$^2$}

{\it $^1$Physics Department,  Brandeis University, Waltham, MA 02454 and \\
Lauritsen Laboratory, California Institute of Technology, Pasadena, CA 91125 \\
{\tt deser@brandeis.edu}

$^2$ Reed College, Portland, OR 97202, USA\\

{\tt jfrankli@reed.edu}}

\end{center}

\begin{abstract}
We derive the Kerr solution in a pedagogically transparent way, using physical symmetry and gauge arguments to reduce the candidate metric to just two unknowns. The resulting field equations are then easy to obtain, and solve. Separately, we transform the Kerr metric to Schwarzschild frame to exhibit its limits in that familiar setting.   
 \end{abstract}

\section{Introduction}

The two fundamental asymptotically flat, Schwarzschild (S) and Kerr (K) [1] solutions, of General Relativity, were derived almost half a century apart, due to the latter's complexity--it is still daunting when 
first encountered. Given K's physical importance, our aim is to provide a transparent, physically instructive, derivation. We will use the labor-saving Weyl method that obviates first wading through the full array of Einstein's equations, then inserting the desired special features of the candidate solution. Instead, we first specify the metric as extensively as possible, using physical and symmetry arguments, then get just the two relevant field equations from the correspondingly reduced Einstein action. This procedure is perfectly legitimate [2], despite appearances. Of course, the equations must still be solved; fortunately they are quite easy. While the metric ansatz, and its plausibility, of course stand on the shoulders of K, the process provides useful insight into its physics.

We will begin by reviewing the oblate spheroidal (OS) coordinates first introduced in this context by [3], then narrow to our candidate metric in this frame. The mechanics of obtaining, and solving, the reduced field equations follows. Separately, we will exhibit the S metric in OS, and that of K in ordinary S, frames. The latter in particular allows one to get a different perspective on K and its limit to S than in OS.

\section{Derivation}
In OS coordinates, the optimal framework for axial symmetry, K has the form [3]
 \begin{eqnarray}\label{KerrBL}
ds^2 &= & - dt^2 + \Sigma  \, dr^2/\Delta  + \Sigma\, d\theta^2 + \of{r^2 + a^2} \, \sin^2\theta \, d\phi^2  + 2 \, m \, r/\Sigma \, \of{dt - a \, \sin^2\theta \, d\phi}^2 ; \\
\Delta &\equiv& r^2 - 2 \, m \, r + a^2, \hskip 1 cm \Sigma \equiv r^2 + a^2 \, \cos^2\theta, \nonumber
\end{eqnarray}
where $(a, m)$ are (the only) constant parameters. [These coordinates are not to be confused with their Cartesian namesakes; as always, they are defined by the interval's form and by their ranges, though the latter are the usual ones.] Deriving this metric from a simple ansatz will be our end-product.
First, consider some limits for orientation. Clearly $m=0=a$ represents flat space in spherical coordinates. The parameter $m$ is aptly named, being the usual ADM mass of the solution (1), that is the system's total energy as measured by an observer at spatial infinity. This is obvious since there (1) coincides with the familiar asymptotic form of S, 
\begin{equation}
ds^2 \sim -\of{1 - 2 m/r} \, dt^2 + (1+2 m/r) \, dr^2 + r^2 \, \of{d\theta^2 + \sin^2\theta \, d\phi^2}.
\end{equation}
We know, by the positive energy theorems of GR, that the vanishing of this single parameter is sufficient 
to imply flat spacetime. Indeed, at $m=0$, (1) is the direct product of time with the textbook metric for Cartesian three-space expressed in OS coordinates. We must next face K's off-diagonal components, with their linear rather than quadratic dependence on $a$, and attendant loss of reflection symmetry. Fortunately, (sub-)intervals of this type are familiar already in flat space, describing rotating systems with angular velocity $a$, and total ADM angular momentum $J=\pm a\, m$. Note finally that the remaining, angular elements $(d \theta^2, d \phi^2)$ retain their flat space (OS) forms, precisely as they do in the S solution in S coordinates: these frames are in fact defined as keeping ``flat" surface area. The differences between K and flat space, then, are entirely contained in the coefficients of $dr^2$ and $(dt - a\, \sin^2 \theta d\phi)^2$, just as they are (without the extra rotation term) for S. [It is, however, surprising that both coefficients in~\refeq{KerrBL} depend only on $r$ and not on $\theta$, as would be expected {\it a priori}.]

The above de-construction of the K metric provides our basis for its re-construction, starting with a metric ansatz with maximal physical and gauge information and minimal number of unknown functions. We propose 
\begin{eqnarray}
ds^2 &= & - dt^2 + \Sigma  \, dr^2/D  + \Sigma\, d\theta^2 + \of{r^2 + a^2} \, \sin^2\theta \, d\phi^2  + Z/\Sigma \, \of{dt - a \, \sin^2\theta \, d\phi}^2 \label{Kansatz} \\
 &=&-dt^2 +   \Sigma \, dr^2/D + \Sigma \,  \of{4 \, q\,  (1 - q)}^{-1} \, dq^2  +  (r^2 + a^2) \, (1-q) \, d\phi^2   +Z/\Sigma \, \of{dt - a \, (1 - q) \, d\phi}^2, \nonumber
\end{eqnarray}
where we have replaced the angle $\theta$ by $q= \cos^2\theta$ for notational convenience.  The kinematical OS factor $\Sigma$ is defined in~\refeq{KerrBL}, and $(Z, D)$ are the two unknowns, to be determined by the field equations. We follow the (simpler) construction of S, in the spirit of [4], keeping $\Sigma$ in $g_{rr}$ (and its reciprocal in $dt^2$), but modify it by the unknown function $D$. Still following the example of S, the coefficient of the whole rotation term, including $dt^2$, involves a different unknown, $Z$. [Time-independence, the hallmark of K, is also assumed,although it might be derivable by the methods of [4].]  Pursuing the analogy still further, we take both unknowns to depend only on $r$; this assumption greatly simplifies the derivation by turning the field equations into ordinary differential ones\footnote{Allowing $(Z,D)$ to depend on $q$ greatly complicates the field equations, contrary to our pedagogical motivation; the more defensible option, letting just $Z$ be general is not too complicated. Note that (as always, [5]) our variables' $q$-independence is only to be invoked in the field equations, after varying the action; however, it is safe to assume time-independence {\it ab initio}.}. While one may argue that introduction of spin should not affect the spherical symmetry of the $dt^2$ coefficient, this is not as defensible here as for S. Rather, its virtue is pragmatic: it is too simple a guess not to be tried first, even if we didn't know it would work.

 Evaluating, and varying, the Einstein action with respect to $(Z,D)$ is best done using an algebraic program, unlike for S, whose explicit action is simple; some irreducible, if purely mechanical, complexity remains. The resulting two equations are  
 
 \begin{eqnarray}
\frac{\delta S}{\delta D} &=& r^2 \, \sqof{D \, \of{Z + r \, (r - Z')} - r^2 \, \of{a^2 + r^2 - Z}} \label{dsdD} \\
&-& a^2 \, \sqof{ r \, D \, (Z'-r) + (r^2 - Z) \, (a^2 + r^2 - Z)} \, q = 0 \nonumber \\
\frac{\delta S}{\delta Z} &=& r^2 \, \sqof{D \, (Z - r^2) - r \, \of{a^2 + r^2 - Z} \, \of{r -D'}} \label{dsdZ} \\
&+& a^2 \, \sqof{D \, \of{2 \, a^2 + r^2 - 2 \, Z} - \of{a^2 + r^2 - Z} \, \of{2 \, a^2 + 3 \, r^2 - 3 \, Z - r \, D'}} \, q = 0 \nonumber
\end{eqnarray}
It is easy to solve these two sets of equations (note that each requires separate vanishing of the $q^0$ and $q^1$ coefficients). Either set is just an ordinary first order equation for $D'$ or $Z'$ , an artifact of the OS system similar to that in S frame, and specifies the correct answer. For example, from~\refeq{dsdD} we learn
\begin{eqnarray}
D &=& \frac{r^2 \, \of{a^2 + r^2 - Z}}{r^2 + Z - r\, Z'} \\
0 &=& \of{a^2 + r^2 - Z}^2 \, Z \, \of{Z - r \, Z'}
\end{eqnarray}
Of the $3$ possible $Z$-roots, only $Z= 2\, m\, r$ (seting the integration constant to agree with that of~\refeq{KerrBL}) is acceptable, as either $Z=0$ or $Z=r^2+a^2=0=D$ lead to singular metrics; this in turn means $D = r^2+a^2- 2\, m\, r$, which is just (1). Likewise, the two equations in~\refeq{dsdZ} are equivalent to the algebraic $D =r^2+a^2- Z$,  
plus an easily solved differential one for $D$ reproducing the result of~\refeq{dsdD}, again modulo 
a singular choice, $D=0$.

This completes the derivation of K from our ansatz~\refeq{Kansatz}. We emphasize that the theorems of [2] guarantee the validity of the Weyl procedure, neither missing any correct solutions nor introducing any spurious ones. 

\section{Schwarzschild in Oblate Spheroidal;  Kerr in Spherical}
In this section, we record the forms (which do not seem available elsewhere) of the S and K solutions in their ``reciprocal" frames: S in OS and K in S. This will shed light on the effects of coordinate choices and more important, permit a better understanding of their relations, especially of S as the non-rotating limit of K. Inserting the spherical(barred)-OS(unbarred) coordinate relations 
\begin{equation}\label{barstoun}
\bar r^2 = r^2 + a^2 \, \sin^2\theta\hskip 1 cm 
\bar \theta =  \hbox{arctan}\of{\frac{\sqrt{r^2 + a^2} \, \tan\theta}{r}} \hskip 1 cm \bar \phi  = \phi.
\end{equation}
into the usual S-interval in S-frame, and using our $q=\cos^2 \theta$ instead of $\theta$, we find for S in OS the form
\begin{eqnarray}\label{SinOS}
ds^2&=& -\of{1 - \frac{2 \, m}{\bar r}} \, dt^2 + \sqof{\frac{r^2/\bar r^2}{1 - \frac{2 \, m}{\bar r}} + \frac{a^4 \,(1-q) \, q}{\of{a^2 + r^2} \, \bar r^2} } \, dr^2  + 2 \,  \frac{a^2 \, m \, r}{\bar r^2 \, \of{2 \, m - \bar r}} \, dr \, dq \\
&+& \sqof{ \frac{r^2 \, (a^2 + r^2) \, (2 \, m -\bar r) - a^4 \, (1 - q) \, q \, \bar r}{4 \, (1 - q) \, q \, \bar r^2 \, (2 \, m - \bar r)} } \, dq^2 +  \of{a^2 + r^2} \, (1-q) \, d\phi^2  \nonumber
\end{eqnarray}
here $\bar r$ stands for its value in~\refeq{barstoun}.  As a check on the correctness of this transformation, we have verified that $R_{\mu\nu}=0$ and that the Riemann invariant matches its usual S value:
\begin{equation}
R_{\alpha\beta\mu\nu} \, R^{\alpha\beta\mu\nu}  = \frac{48 \, M^2}{\bar r^6}. 
\end{equation}
Note that, surprisingly, the S metric has now acquired an-off diagonal, $dr\, dq$, component. Indeed, were we to ansatz a diagonal metric in OS coordinates for S, the resulting field equations would force us back to S-frame\footnote{While the $2$-dimensional $(r, q)$ subspace can of course always be diagonalized, that would takes us outside the OS frame. }, by stating that the only such solutions require $a=0$.  It is clear that the $a=0$ limit of~\refeq{SinOS} is the initial S, while $m=0$ is indeed flat space: $g_{rq}=0$ and $(g_{rr} , g_{qq})$ limit to their flat OS forms.

Next, we transform K to S-frame, which is better suited to understanding the S-limit of K, as we now see. The required coordinate relation, inverse to~\refeq{barstoun}, is
\begin{equation}\label{rqrbqb}
2\, \left\{ r^2, a^2 \, q \right\} =  \sqrt{a^4 - 2 \, a^2 \, (1 - 2 \, \bar q) \, \bar r^2 + \bar r^4} + (\bar r^2 - a^2) \, \left\{ 1 , -1\right\}
\end{equation}
(where $\bar q = \cos^2\bar\theta$ for the spherical $\bar\theta$) with $(t, \phi)$ unchanged. We then obtain K as
\begin{eqnarray}\label{KDSS}
ds^2 &=& -\of{1 - \frac{2 \, m \, r}{\rho^2}}  \, dt^2  - \frac{4 \, a \, m \, r (1-q)}{\rho^2} \, dt \, d\phi +  \frac{r^2 \, (r^2 + a^2)^2 + a^4 \, (1 - q) \, q \, \Delta}{\Delta\, \bar r^2 \, \rho^2}  \, dr^2 \\
&+& \frac{2 \, a^2 \, m \, r \, \bar r}{\Delta \, \rho^2} \, dr \, dq + \frac{\of{ a^4 \, (1 - q) \, q + r^2 \, \Delta}\, \bar r^4}{4 \, (1 - q) \, q \, r^2 \, \Delta}  \, dq^2 +  \frac{(1 - q) \, \of{2 \, a^2 \, m \, r \, (1-q) + (r^2 + a^2) \, \rho^2}}{\rho^2}  \, d\phi^2 \nonumber \\
\rho^2 &\equiv& r^2 + a^2 \, q \nonumber
\end{eqnarray}
keeping the unbarred coordinates for notational simplicity; they merely mean their values in terms of the  barred ones given in~\refeq{rqrbqb}. Note the appearance of the off-diagonal $dr\, dq$ in addition to the spinning part. The reason it is more illuminating to take the K$\longrightarrow$  S limit here than in OS is that one wants to shut off the angular momentum $\sim a\, m$, while keeping $m$ finite. But this forces setting $a=0$, which is overkill in OS, involuntarily also taking us from OS to just S. Instead, since the K of~\refeq{KDSS} is already in S-frame, the parameter $a$ no longer refers to the OS kinematics, but more properly just to the rotation itself. It is clear that the barred and unbarred coordinates coincide in this limit, so that~\refeq{KDSS} reduces to the elementary S. Expansion of K$=$S $+0(a) +0(a^2)$ can then be used to understand the cumulative effects of rotation on the ``base" S, beyond the simple linearized inclusion of $J=a\, m$, an exercise we omit here.

\section{Summary}
We have derived the Kerr [1] solution using physical--axial symmetry and constant rotation--information, 
expressed in the OS [3] gauge. The resulting metric ansatz was further reduced by hints from similar derivation of S, and demanding maximal simplicity. Thanks to the power of the Weyl method [2], this led 
directly to just the two field equations for the unknown metric components, which were then easily solved. Separately, we shed additional light on the K$\longrightarrow$ S  limit by first transforming K to S-frame.

We thank Matt Visser for useful correspondence. SD acknowledges support from grants NSF PHY 07-57190 and DOE DE-FG02-164 92ER40701.

\end{document}